\documentclass[useAMS,usenatbib]{mnras}

\usepackage{graphicx}
\usepackage{epstopdf}
\usepackage{amsmath,esint}

\usepackage{accents}
\newcommand*{\dt}[1]{%
   \accentset{\mbox{\large\bfseries .}}{#1}}

\usepackage{etoolbox}
\makeatletter
\patchcmd\@combinedblfloats{\box\@outputbox}{\unvbox\@outputbox}{}{%
   \errmessage{\noexpand\@combinedblfloats could not be patched}%
}%
 \makeatother

\title[Not an Oxymoron: Some XBPs with Enormous $\dt{P_S}$ Reveal Weak $B_*$]{Not an Oxymoron: Some X-ray Binary Pulsars with Enormous Spinup Rates Reveal Weak Magnetic Fields}

\author[D. M. Christodoulou et al.]{D. M. Christodoulou,$^{1,2}$ S. G. T. Laycock,$^{1,3}$ and D. Kazanas$^{4}$
\\
$^{1}$Lowell Center for Space Science and Technology, University of Massachusetts Lowell, Lowell, MA, 01854, USA.\\
$^{2}$Dept. of Mathematical Sciences, Univ. of Massachusetts Lowell, 
Lowell, MA, 01854, USA. E-mail: dimitris\_christodoulou@uml.edu\\
$^{3}$Dept. of Physics \& Applied Physics, Univ. of Massachusetts Lowell, Lowell, MA, 01854, USA. E-mail: silas\_laycock@uml.edu\\
$^{4}$NASA/GSFC, Laboratory for High-Energy Astrophysics, Code 663, Greenbelt, MD 20771, USA. E-mail: demos.kazanas@nasa.gov
}

\begin{document}

\def\gsim{\mathrel{\raise.5ex\hbox{$>$}\mkern-14mu
                \lower0.6ex\hbox{$\sim$}}}

\def\lsim{\mathrel{\raise.3ex\hbox{$<$}\mkern-14mu
               \lower0.6ex\hbox{$\sim$}}}

\pagerange{\pageref{firstpage}--\pageref{lastpage}} \pubyear{2017}

\maketitle

\label{firstpage}

\begin{abstract}
Three high-mass X-ray binaries have been discovered recently exhibiting enormous spinup rates. Conventional accretion theory predicts extremely high surface dipolar magnetic fields that we believe are unphysical. Instead, we propose quite the opposite scenario: some of these pulsars exhibit weak magnetic fields, so much so that their magnetospheres are crushed by the weight of inflowing matter. The enormous spinup rate is achieved before inflowing matter reaches the pulsar's surface as the penetrating inner disk transfers its excess angular momentum to the receding magnetosphere which, in turn, applies a powerful spinup torque to the pulsar. This mechanism also works in reverse: it spins a pulsar down when the magnetosphere expands beyond corotation and finds itself rotating faster than the accretion disk which then exerts a powerful retarding torque to the magnetic field and to the pulsar itself. The above scenaria cannot be accommodated within the context of neutron-star accretion processes occurring near spin equilibrium, thus they constitute a step toward a new theory of extreme (far from equilibrium) accretion phenomena.
\end{abstract}




\section{Introduction}\label{intro}

\subsection{Standard Accretion Theory and Extremes}

In observations of high-mass X-ray binaries (HMXBs) to date, the results are routinely being interpreted by using the theoretical results of Ghosh and collaborators \citep{gho77,gho79}. We have been guilty of this practice as well. At the same time, we know that this model of accretion is not the definitive word on the subject because the range of validity of this theory is heretofore unspecified. For instance, imagine a HMXB pulsar with an especially weak surface magnetic field ($B\sim 1$~GG). Then the entire theory is obviously not applicable since a modest/large increase of matter inflow $\dt{\cal M}$ during outburst would easily crush the magnetic field back on to the pulsar's surface \citep[see, e.g.,][]{klu15}. On the opposite end \citep[see, e.g., recent work by][]{bri18}, many believe that some HMXB pulsars harbor magnetar-strength magnetic fields ($B\geq 100$~TG) ignoring the fact that such fields would easily push the magnetospheres out in the direction of the light cylinder radius
\begin{equation}
r_{lc} \equiv \frac{c\, P_S}{2\pi}  = 4.77\times 10^9\left(\frac{P_S}{1~{\rm s}}\right)~{\rm cm} \, ,
\label{rlc}
\end{equation}
where $c$ is the speed of light and $P_S$ is the pulsar's spin period;
obliterating in the process the inner region of the disk and shutting off accretion and variability altogether; a feat that so far has only been accomplished by a few millisecond pulsars with relatively strong magnetic fields \citep[$B\sim 0.1$~GG; e.g.,][]{cam98,hartman08,pat09,chr18} in which the light cylinder is not too far beyond the corotation radius
\begin{equation}
r_{co} = 1.68\times 10^8\left(\frac{P_S}{1~{\rm s}}\right)^{2/3}~{\rm cm} \, .
\label{corot}
\end{equation}
For a 1 ms pulsar, the ratio $r_{lc}/r_{co} < 3$.

In HMXBs, the above two extreme cases of pulsar magnetic fields signal the need for new evolutionary paths, and these paths cannot possibly be accommodated in the context of standard accretion theory \citep{fra02}. For strong fields, only a dramatic increase in $\dt{\cal M}$, such as that during outbursts, can compete and push the magnetosphere back inside corotation \citep{chr18}. For weak fields, there is no hope: the magnetosphere will be crushed by the inflow in every single outburst and the process of accretion will change character dramatically \citep{pri72,wan97,car17,car18,gon18}; in particular, the spinup rate will jump to unprecedented levels during prograde accretion. This smoking gun and its connection to ``retrograde accretion'' \citep{chr17a} is the subject of the present work.  

\subsection{Observations of Enormous Spinup Rates}

In \cite{chr17a}, we proposed a new evolutionary path for Be/X-ray pulsars in the $\dt{P_S}$ versus $P_S$ diagram that shows reflection symmetry about the $\dt{P_S}=0$ line at long periods ($P_S>140$~s). In this scenario, long period pulsars with $\dt{P_S}>0$ are expected to reverse their spin evolution and transition rapidly below the $\dt{P_S}=0$ line where they will continue to spin up; and they may even return to much shorter periods, provided that subsequent accretion events continue to be prograde (in the direction of their spins).

The two longest-period pulsars with high-quality archival data are SXP1062 ($\dt{P_S}=+8.2\times 10^{-8} \pm 2.5\times 10^{-8}$) and SXP1323 ($\dt{P_S}=-7.5\times 10^{-8} \pm 3.2\times 10^{-8}$) \citep[dimensionless values converted from those listed in][]{yang17}. Based on these $\dt{P_S}$ values, we surmised that SXP1062 is the best candidate for switching to spinning up in the years following 2014, whereas SXP1323 has already switched in the opposite direction some time prior to year 2000. These expectations turned out to be inaccurate because the \cite{yang17} pipeline does not access all available X-ray observations (especially {\it Suzaku} observations prior to 2010 and any observations beyond early 2014).

In a key paper, \cite{car17} showed that SXP1323 clearly switched in 2005, a result that is also detectable in our pipeline data, albeit with considerable difficulty. This source continued to spin up with $\dt{P_S}=-6.86\times 10^{-7}$ from 2006 to late 2016, the end of the 2.5-10~keV data sequence. This rate is nearly one order of magnitude faster than that determined from our pipeline data, a consequence of the pipeline not including all of the \cite{car17} data.

In another key paper, \cite{gon18} showed that SXP1062 may also have switched in late June 2014, but this result is not confirmed because of yet another reversal in the last two observations. The $\dt{P_S}$ values for the observations of SXP1062 are listed in Table~\ref{table1} in intervals of about 10 days and they are truly enormous. It is interesting to note that these unprecedented rates occur while the 0.3-10~keV luminosity continues to decrease monotonically past maximum in the last two {\it Chandra} observations.

\cite{car18} have just reported an enormous spinup rate, $P_S=-5.56\times 10^{-7}$, over 4 days in the 0.2-20~keV combined range of {\it XMM-Newton} (0.2-10~keV) and {\it NuSTAR} (3-20~keV) for ULX1 in NGC300, an ultraluminous X-ray (ULX) pulsar with $P_S=31.54$~s. This $\dt{P_S}$ value is comparable to that reported by the same group for SXP1323 and it is expected to drive the spin period down to about 20~s in the next few months. Judging from the much smaller spinup rates of the other three known ULX pulsars \citep{bac14,fur16,isr17a,isr17b}, changes in the spin of this fourth ULX pretender may taper off when its spin period declines by another order of magnitude to $P_S\sim$ 1~s.

A large spinup rate of $\dt{P_S}=-1.1\times 10^{-7}$ was also reported for the transient X-ray 18~s pulsar CXOU J073709.1+653544 in NGC 2403 \citep{tru07}. In its only detected ouburst by {\it Chandra} and {\it XMM-Newton}, the 0.3-7~keV luminosity reached a maximum of $1.5 L_{Edd}$, where 
$$
L_{Edd}=1.77\times 10^{38}~{\rm erg~s}^{-1}\, ,
$$ 
for the canonical pulsar mass of $1.4 M_\odot$. Unfortunately, no other observations of this pulsar exist in the {\it Chandra} and {\it XMM-Newton} archives. 
We do not include the limited information about this pulsar in our sample, but its example
serves as a reminder that the above discussed rapid spin evolutions are not just isolated incidents, instead they may be quite common among HMXB pulsars during their type I and type II outbursts \citep{coe10,reig11}.

\subsection{Outline}

SXP1323 has unequivocally switched to spinning up in 2005 \citep{car17} and follow-up observations of NGC300 ULX1 are currently under way \citep{vas18}. This leaves SXP1062 for which we examine the chronology of its 2014 outburst in \S~\ref{1062} using the latest data from \cite{gon18} and data from our pipeline \citep{yang17}. The behavior of this pulsar past outburst cannot be understood in the context of standard accretion theory. We quantify this prominent failure in \S~\ref{std}, where we summarize the observed and calculated properties of the three pulsars introduced above. 

In \S~\ref{new_path}, we describe a new evolutionary path for such pulsars that explains the enormous spinup rates without requiring absurd values of their propeller-line luminosities and their magnetic fields and, surprisingly, it also sheds light to the process of spindown during prograde and retrograde accretion. This is because, unlike standard accretion theory, the new mechanism can operate in retrograde accretion disks to some limited extent.  In \S~\ref{conc}, we summarize our conclusions.

\begin{table*}
\caption{Observations of SXP1062$^{(a)}$}
\label{table1}
\begin{tabular}{cccccc}
\hline
Date & MJD & $P_S$ & $L_X$  & $\Delta P_S/\Delta t$ & $\Delta(\ln L_X)/\Delta t$ \\
 & & (s) & (erg~s$^{-1}$) & (s s$^{-1}$) & (\%) \\
\hline
2010-03-25 & 55280.00	&1062.00	& $6.3\times 10^{35}$ &  &  \\
2012-10-14 & 56214.00	&1071.01	& $2.6\times 10^{36}$ & $1.1\times 10^{-7}$ & 0.33 \\
2013-10-11 & 56606.80	&1077.97	& $5.7\times 10^{35}$ & $2.1\times 10^{-7}$ &  $-0.20$\\
2014-06-19 & 56827.80	&1091.10	& $2.6\times 10^{36}$ & $6.9\times 10^{-7}$ &  1.6\\
2014-06-29 & 56837.49	&1087.10	& $3.0\times 10^{36}$ & $-4.8\times 10^{-6}$ &  1.6\\
2014-07-08 & 56846.30	&1079.30	& $2.4\times 10^{36}$ & $-1.0\times 10^{-5}$ &  $-2.3$\\
2014-07-18 & 56856.91	&1086.00	& $1.6\times 10^{36}$ & $7.3\times 10^{-6}$ &  $-3.1$\\
\hline
\end{tabular}
\\
(a)~Data from \cite{gon18} and references therein.
\end{table*}

\begin{table*}
\caption{HMXB Pulsars with Enormous Spinup Rates$^{(a)}$}
\label{table2}
\begin{tabular}{lccccccccc}
\hline
Pulsar & $P_S$ & $\dt{P_S}$ & $L_{max}$ & $b$ & $L_{p2}/b$ & $B_2$ & $L_{min}$ & $L_{p1}/b$ & $B_1$ \\
 & (s) & (s\,s$^{-1}$) & (erg~s$^{-1}$) & & (erg~s$^{-1}$) & (TG) & (erg~s$^{-1}$) & (erg~s$^{-1}$) & (TG) \\
\hline
SXP1323 & 1100  & $-6.86\times 10^{-7}$  & $1.7\times 10^{36}$\,$^{(b)}$ & 1 & $1.0\times 10^{35}$  & 252  & $4.3\times 10^{34}$\,$^{(b)}$  & $4.3\times 10^{34}$  & 164  \\
SXP1062 & 1079  & $-2.03\times 10^{-6}$\,$^{(c)}$  & $3.0\times 10^{36}$ & 1 & $3.2\times 10^{35}$  & 434  & $5.7\times 10^{35}$  & $5.7\times 10^{35}$  & 583  \\
N300 ULX1 & 31.54  & $-5.56\times 10^{-7}$  & $4.4\times 10^{39}$ & 25 & $1.3\times 10^{37}$  & 45.6  & $4.0\times 10^{36}$\,$^{(d)}$  & $1.6\times 10^{35}$  & 5.01  \\
\hline
\end{tabular}
\\
(a)~Data from \cite{car17}, \cite{gon18}, and \cite{car18}, in rows 1-3, respectively. (b)~Calculated for a distance to SMC of 60~kpc. (c)~Average value over the last four observations in Table~\ref{table1} (29 days). (d)~Data from \cite{bin16}.
\end{table*}

\section{The 2014 Outburst of SXP1062}\label{1062}

After its {\it XMM-Newton/Chandra} discovery \citep{hen12}, SXP1062 was observed again by {\it XMM-Newton} \citep{hab12,stu13} and a value of $\dt{P_S}=+7.2\times 10^{-8} \pm 1.4\times 10^{-8}$ was measured. This rate agrees with our pipeline value to within the quoted uncertainties. The source was revisited by {\it XMM-Newton/Chandra} in the period 2013-14 \citep{gon18}. The combined data set that includes past observations \citep{hen12,stu13} along with the new observations (listed here in Table~\ref{table1}) indicates that SXP1062 switched its accretion mode a few days after June 19, 2014, except for the last two data points in which the trend appears to have reversed once again.

The \cite{gon18} data show initially $\dt{P_S}=+1.9\times 10^{-7}$ (from linear regression of the earliest four observations in Table~\ref{table1}, years 2010-14), about 2.5 times faster than previous measurements \citep{stu13,yang17}; they also show that the spin period derivative switched to an average value of $\dt{P_S}=-2.03\times 10^{-6}$ during subsequent evolution.

But in the last observation taken in July 18, 2014 the source switched back (and showed an astonishing value of $\dt{P_S}=+7.3\times 10^{-6}$), just as the type I outburst was powering down: Between the last three observations separated by 9-10 days from one another, $P_S$ decreased by $7.8$~s ($-0.7$\%) during the first 9 days and then increased by $6.7$~s ($+0.6$\%), while the X-ray luminosity $L_X$ kept decreasing systematically throughout by 2-3\% per day for 19 days (Table~\ref{table1}). 

Such a tremendous bounce in $P_S$ values cannot be understood physically in the framework of standard accretion theory \citep{ill75,gho77,gho79,ste86,fra02} because the mass flow rate $\dt{\cal M}$ to the pulsar is decreasing monotonically as the outburst is powering down; thus, the $\dt{P_S}$ of the accreted matter can change abruptly neither sign nor magnitude by the observed factor of $(+7.3\times 10^{-6})/(-2.03\times 10^{-6})=-3.6$. These puzzling results find a natural explanation in terms of a new evolutionary path that we describe in \S~\ref{new_path} below.

\section{Results from Standard Accretion Theory}\label{std}

We summarize in Table~\ref{table2} the observed properties of the three pulsars introduced above and the results of calculations using standard accretion theory. The observed values of $P_S$, $\dt{P_S}$, $L_{max}$, and $L_{min}$ are inputs to the calculations. 

If $L_{max} >> L_{Edd}$, we assume anisotropic emission with a beaming factor of \citep{chr17b}
\begin{equation}
b = \frac{L_{max}}{L_{Edd}} > 1\, .
\label{beaming}
\end{equation}
Beaming turns out to be important only for NGC300 ULX1 and it causes a decrease of the propeller-line luminosity and the magnetic field. Dropping this assumption for NGC300 ULX1 only makes the results listed in Table~\ref{table2} seem even more absurd.

The propeller-line luminosity is determined in two different ways and then it is scaled by the beaming factor $b$: 

First, we assume that $L_{p1} = L_{min}$, where $L_{min}$ is the lowest luminosity observed in quiescence or as far below the peak of the outburst $L_{max}$. Then, the surface dipolar magnetic field $B_1$ is determined from the equation \citep{ste86}
\begin{equation}
L_{p} = 2\times 10^{37} \left(\frac{\mu_*}{10^{30}~{\rm G~cm^3}}\right)^2 
\left(\frac{P_S}{1~{\rm s}}\right)^{-7/3} ~{\rm erg~s^{-1}}\, ,
\label{stella}
\end{equation}
for $L_{p}=L_{p1}/b$, where canonical pulsar parameters have been used (mass $M_*=1.4 M_\odot$ and radius $R_*=10$~km) and the magnetic moment is defined by
\begin{equation}
\mu_{*}\equiv B_{1} R_*^3 \, . 
\label{mu}
\end{equation}
Eq.~(\ref{stella}) does not depend on $\dt{P_S}$, thus it does not rely on torque balance at the inner edge of the accretion disk, but it assumes that the magnetospheric radius coincides with the corotation radius. As such, this equation includes no dependency on the fastness parameter or torque efficiency \citep{gho77,gho79,wan97}, which certainly makes it uncertain. But in practice, it has worked quite well \citep{chr16,chr18}.

Second, we use $P_S$ and $\dt{P_S}$ in the equation 
\begin{equation}
L_{p2} = \frac{\eta}{2}\left(2\pi I_* |\dt{P_S}|\right)\left( \frac{2\pi}{P_S^7}\frac{G M_*}{R_*^3} \right)^{1/3} \, ,
\label{lmin2}
\end{equation}
\citep[for details, see][]{chr17b}, where $\eta = 0.5$ is the efficiency of converting half of the accretion power to X-rays, $I_*=2M_*R_*^2/5$, and $G$ is the gravitational constant. This equation shows that $L_{p2}\propto |\dt{P_S}|$ and it naturally produces extremely high values of $B_2\propto |\dt{P_S}|^{1/2}$ when $L_p=L_{p2}/b$ is used into eq.~(\ref{stella}).

The calculated results listed in Table~\ref{table2} are unreasonable. The Magellanic pulsars whose outbursts did not even come close to the level of the Eddington rate appear to have magnetic fields substantially above the quantum limit 
$$
B_{QL} = 44.14~{\rm TG}\, ,
$$
which is unacceptable in our opinion; and the two determinations of the magnetic field of the ULX pulsar are in strong disagreement.\footnote{The observed $L_{min}$ values may not reflect the correct propeller-line luminosities, yet they cannot be raised; they can only be lowered by future fainter detections. But that would only exacerbate the discrepancies in the determinations of the magnetic fields.}

Thus, it becomes obvious that the enormous values of $\dt{P_S}$ cannot be used in conjunction with standard theory to predict  the basic properties of these pulsars. If we assume that the observed $\dt{P_S}$ values are not in error, then we must conclude that the above standard equations are not applicable to accretion states such as these studied in this work.

\section{A New Evolutionary Path}\label{new_path}

\subsection{Beyond Spin-Equilibrium Theory}\label{beyond}

All the problems that we encountered above disappear if we assume that the magnetic fields of these pulsars are weak and the mass inflow during outbursts can crush the magnetospheres \citep{pri72}. In this case, new phenomena are expected to occur that cannot be described by the equations used in \S~\ref{std}. The main difference is that the inner disk radius $R_d$ needs to be determined strictly from observations, not from theoretical assumptions as to the size and behavior of the magnetosphere.

\begin{table*}
\caption{Properties of Fully Compressed Magnetospheres}
\label{table3}
\begin{tabular}{lccccccc}
\hline
Pulsar & $P_S$ & $\dt{P_S}$ & $L_{max}$ & $r_{co}$ & $R_d=r_{mag,min}$ & $R_{d}/r_{co}$ & $B_*$ \\
          & (s) & (s\,s$^{-1}$) & (erg\,s$^{-1}$) & (cm) & (cm) &  & (TG) \\
\hline
SXP1323            & 1100      & $-6.86\times 10^{-7}$ & $1.7\times 10^{36}$ & $1.8\times 10^{10}$ & $6.4\times 10^7$  & $3.6\times 10^{-3}$   &  0.03 \\
SXP1062            &  1087.1   & $-4.80\times 10^{-6}$ & $3.0\times 10^{36}$ & $1.8\times 10^{10}$ & $1.1\times 10^9$  & $6.0\times 10^{-2}$ &  6 \\
N300 ULX1         &  31.54    & $-5.56\times 10^{-7}$ & $4.4\times 10^{39~(a)}$ & $1.7\times 10^9$ & $9.3\times 10^6$  & $5.6\times 10^{-3}$  &  0.06 \\
\hline
\end{tabular}
\\
(a)~Isotropic X-ray luminosity. Beaming is discussed in the text.
\end{table*}

In this mechanism, the higher $\dt{\cal M}$ during outburst drives the accretion disk inside the corotation radius and compresses the magnetosphere to a much smaller radius $R_d<< r_{co}$. Then the inner disk finds itself rotating faster than the pulsar and its magnetosphere, thus it torques the magnetic field lines forward which, in turn, drag the pulsar forward producing an enormous negative $\dt{P_S}$ before any matter gets loaded on to the lines and descends toward the pulsar's surface. Thus, the pulsar spins faster as a result of ``action at a distance'' (not by an increase of the specific angular momentum in its accretion column); and the observed $\dt{P_S}$ does not reflect the observed X-ray luminosity, thus eq.~(\ref{lmin2}) is invalid. 

The inner disk radius $R_d$ is determined by torque balance and observed quantities ($P_S$, $|\dt{P_S}|$, and $L_{X}$). On the equatorial plane of the disk (assumed to be an ADAF disk), $R_d$ coincides with the spherical magnetospheric radius $r_{mag}$. As was also derived by \cite{klu15} (their eq.~(8)), we find that
\begin{equation}
R_d = r_{mag} = \left(\frac{2\pi\eta}{5}\right)^2 \left(\frac{G M_*^3 R_*^2}{\left(L_{X}/b\right)^2}\right)  \left(\frac{\dt{P_S}^2}{\left.P_S\right.^4}\right) \, ,
\label{rmag}
\end{equation}
where the beaming factor $b\geq 1$, the efficiency $\eta = 1/2$ \citep{chr17b}, and canonical pulsar values are to be used for $M_*$ and $R_*$. In what follows, we assume strictly isotropic X-ray emission (as observers routinely do) and we set $b=1$. \cite{klu07} reported a range of 2-3 for $r_{mag}$ from a literature search, but this also includes cases where this radius was determined by pressure balance. In our case, the error in $r_{mag}$ depends only on the efficiency of angular momentum transfer from the disk to the neutron star and, as such, it is expected to be comparable to the same factor of 2-3. 

Eq.~(\ref{stella}) cannot be used in the determination of the magnetic field $B_*$ because it assumes that $r_{mag}=r_{co}$. But $B_*$ can be determined from pressure balance at the interface $R_d=r_{mag}$ during full magnetospheric compression ($L_X = L_{max}$). We find that
\begin{equation}
(B_*)^4 = \left(\frac{\eta}{2}\right)^{-2} \left[\frac{(L_{max})^2\, (r_{mag,min})^7}{G M_* R_*^{10}}\right] \, ,
\label{B1}
\end{equation}
where $r_{mag,min}$ must be determined from eq.~(\ref{rmag}) for $L_X=L_{max}$. Using eqs.~(\ref{rmag}) and~(\ref{B1}), we determine the values of $r_{mag,min}$ and $B_*$ for the pulsars listed in Table~\ref{table2}. Our results are collected in Table~\ref{table3}. The magnetic field values in Table~\ref{table3} are highly uncertain because of the steep dependence of $r_{mag,min}$ on the observed quantities ($P_S^{-4}$, $\dt{P_S}^2$, and $(L_{max})^{-2}$). The only safe conclusion appears to be that the magnetic fields of SXP1323 is very weak (below the lowest Magellanic propeller line with $B_*=0.29$~TG), whereas the magnetic field of SXP1062 is surprisingly strong.

For NGC300 ULX1, the isotropic X-ray luminosity is listed in Table~\ref{table3} and then $B_*$ appears to be very weak. But these results do not hold in the case of beaming. If $L_{max}$ is reduced by a beaming factor of 25 (Table~\ref{table2}), then $R_d/r_{co}=3.44$ which is obviously wrong. If we assume instead that $L_{max}=4-5\, L_{Edd}$, then we obtain, respectively, $R_d/r_{co}=0.215-0.138$ and $B_*=13.6-6.95$~TG. Reasonable as they may be, these values are in strong disagreement with the isotropic values listed in Table~\ref{table3}.

\begin{table*}
\caption{Movement of the Magnetosphere of SXP1062 During its 2014 Outburst$^{(a)}$}
\label{table4}
\begin{tabular}{cccccc}
\hline
Date & $d^{(b)}$ & $P_S$ & $L_X$  & $\Delta P_S/\Delta t$ & $r_{mag}/r_{co}$ \\
 & & (s) & (erg~s$^{-1}$) & (s s$^{-1}$) &  \\
\hline
2012-10-14 & 0	&1071.01	& $2.6\times 10^{36}$ & $^{(c)}$ &  \\
2013-10-11 & 392.80	&1077.97	& $5.7\times 10^{35}$ & $2.1\times 10^{-7}$ &  $3.1\times 10^{-3}$ \\
2014-06-19 & 613.80	&1091.10	& $2.6\times 10^{36}$ & $6.9\times 10^{-7}$ &  $1.6\times 10^{-3}$ \\
2014-06-29 & 623.49	&1087.10	& $3.0\times 10^{36}$ & $-4.8\times 10^{-6}$ &  $6.0\times 10^{-2}$ \\
2014-07-08 & 632.30	&1079.30	& $2.4\times 10^{36}$ & $-1.0\times 10^{-5}$ &  $4.4\times 10^{-1}$ \\
2014-07-18 & 642.91	&1086.00	& $1.6\times 10^{36}$ & $7.3\times 10^{-6}$ &  $4.9\times 10^{-1}$ \\
\hline
\end{tabular}
\\
(a)~Data from \cite{gon18} and references therein. (b)~$d\equiv$ MJD $-$ 56214. (c)~This forward difference of $\Delta P_S/\Delta t$ is not meaningful because the nearest data point before day $d=0$ is 2.5 years in the past (Table~\ref{table1}). For the same reason, the next two entries shown here are also questionable.
\end{table*}

\begin{figure}
\begin{center}
    \leavevmode
      \includegraphics[trim=0 0cm 0 0.1cm, clip, angle=0,width=9 cm]{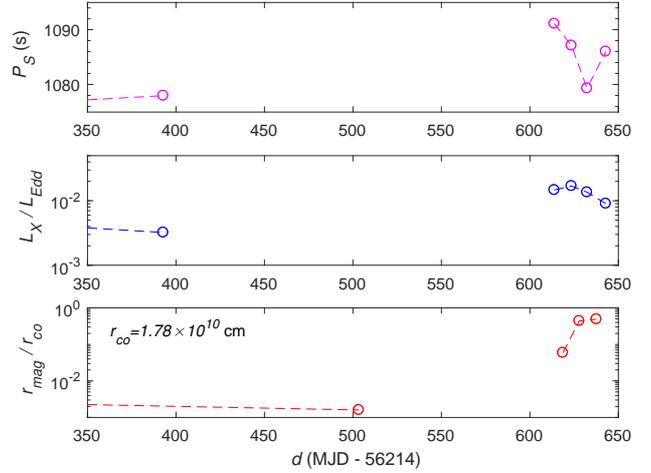}
\caption{Spin period, luminosity, and magnetospheric radius variation during the evolution of SXP1062 \citep[based on the data in][and in Table~\ref{table4}]{gon18}. Inflow pushes inward but the magnetic field resists and eventually pushes outwad in a continual tug of war that is described in the text. In the bottom panel, the $r_{mag}/r_{co}$ values have been shifted to the midpoints of the dates of observations.
\label{fig1}}
  \end{center}
\end{figure}

The magnetosphere can no longer be squeezed when $\dt{\cal M}$ drops substantially past outburst. It will then push back out, leading to episodic variability such as that observed from the above pulsars. Upon further expansion, the field lines will find themselves rotating faster than the disk and the interaction will torque the pulsar in the retrograde direction, slowing down its rotation. This phenomenon is evident in SXP1062 (\S~\ref{1062}, Table~\ref{table1}, and Fig.~\ref{fig1}). When $\dt{\cal M}$ dropped to about 50\% of maximum, this pulsar reversed and showed an enormous spindown rate of $\dt{P_S} = +7.3\times 10^{-6}$. More observations are needed to determine whether this is a temporary bounce or the magnetosphere will keep pushing outward.
 
Using eqs.~(\ref{rmag}) and~(\ref{corot}), we show in Table~\ref{table4} and in Fig.~\ref{fig1} the movement of the magnetosphere of SXP1062 before/after the source reached its maximum X-ray luminosity. Prior to day 393, all quantities appear to be nearly flat. There are no observations in the interval of $d=393$-614 days, but the magnetosphere appears to be fully compressed, at least in a part of this interval. We do not trust this behavior because of the lack of observations. At some time prior to $d=614$, the magnetic field must have expanded,\footnote{We suspect that the magnetosphere expanded prior to $d=586$ and then it was pushed back during $d = 586$-621 (Fig.~\ref{fig1}) because a strong spinup glitch was reported by \cite{ser17} in this interval. Their result fits so well in the \cite{gon18} time line (Table~\ref{table4}) that we also suspect that future glitches in HMXBs will be indicative of jerking of their magnetospheres.}
and this is why we do not see its compression in the first instance that the 2014 outburst was observed. We see however a sharp decrease of $P_S$ and an expansion of $r_{mag}$ as soon as $L_X$ declines past maximum. This expansion is responsible for the bounce in $P_S$ seen in the last two points. The fact that the magnetic field can cause such a spindown tells us that $B_*$ is strong and Table~\ref{table3} lists our estimate of $B_*=6$~TG, one of the highest values ever found among Magellanic HMXB pulsars.

The scenario that we just described for SXP1062 relies on sparse data. It will have to be confronted by denser data sets (e.g., daily observations) of HMXBs exhibiting high values of $|\dt{P_S}|$ in outburst. Observers need to be aware that an expanding magnetosphere in quiescence can easily be missed (as in Fig.~\ref{fig1}) because major X-ray bursts are usually caught several days late. But the expansion that occurs past maximum luminosity when an outburst powers down should always be seen.

\subsection{The Physics of Extreme Accretion}\label{phys}

The equations in \S~\ref{beyond} are convenient for calculations but too opaque for physical interpretations. We rewrite eq.~(\ref{rmag}) in physical form, viz.
\begin{equation}
R_{d} = r_{mag} = R_*\left(\frac{I_* \dt{\Omega}_S}{\dt{\cal M} R_* V_{K*}}\right)^2 \, ,
\label{rmag2}
\end{equation}
where $I_*$ is the canonical moment of inertia of the pulsar, $\Omega_S=2\pi/P_S$, and
\begin{equation}
V_{K*} = \sqrt{G M_*/R_*} \, .
\label{vk}
\end{equation}
Interpretation of eq.~(\ref{rmag2}) is now straightforward: the numerator of the fraction represents $\dt{\cal L}_p$, the rate of angular momentum change of the pulsar, whereas the denominator represents a scaled version of $\dt{\cal L}_d$, the rate of angular momentum transferred from the disk. If the disk reaches all the way to the surface of the star, these two rates are equal and then $R_d=r_{mag}=R_*$. But if the disk's $\dt{\cal M}$ originates at a large distance where the disk is truncated by the magnetosphere, then naturally $R_d=r_{mag} > R_*$.

Next we rewrite eq.~(\ref{B1}) in physical form, viz.
\begin{equation}
(B_*)^4 = \left(\frac{V_{K*}\dt{\cal M}_{max}}{R_*^2}\right)^2 \left(\frac{r_{mag,min}}{R_*}\right)^7 \, ,
\label{B2}
\end{equation}
where $\dt{\cal M}_{max}$ and $r_{mag,min}$ are calculated for $L_X=L_{max}$. The additional factor of $(\dt{\cal M}_{max})^2$ in the first parenthesis simply rescales the power of  the $\dt{\cal M}_{max}$ hidden in the ratio $(r_{mag,min}/R_*)^7$ from $(\dt{\cal M}_{max})^{-14}$ to $(\dt{\cal M}_{max})^{-12}$. The various starred quantities in eq.~(\ref{B2}) rescale the terms and produce the correct units, so this equation simply shows that $B_*^4$ is proportional to $(\dt{\Omega}_S)^{14}/(\dt{\cal M}_{max})^{12}$ or equivalently, that
\begin{equation}
B_* \, \propto \, |\dt{\Omega}_S|^{7/2} \, |\dt{\cal M}_{max}|^{-3} \, .
\label{scale}
\end{equation}
This scaling shows that a large $|\dt{\Omega}_S|$ may indicate a stronger field that can transfer efficiently the disk torque to the pulsar; but a large $|\dt{\cal M}_{max}|$ may indicate a weaker field that can be compressed easier by the inflow. The two processes compete against one another and this is why observations of extremely high $|\dt{\Omega}_S|$ values \citep{car17,car18,gon18}, without knowledge of $\dt{\cal M}_{max}$, do not imply automatically a strong magnetic field. In fact, the powers of the two competing quantities are not too dissimilar, thus for all practical purposes, the strength of $B_*$ is decided by the ratio $|\dt{\Omega}_S/\dt{\cal M}_{max}|$, where both quantities must be determined at maximum X-ray luminosity. 

\subsubsection{Maximum Mass Inflow Rate During Outburst}

On the other hand, eq.~(\ref{scale}) for a fixed value of $B_*$ indicates that
$$
|\dt{\Omega}_S|\propto (\dt{\cal M}_{max})^{6/7}\, ;
$$ 
this makes sense since a larger spinup rate does always result from a larger inflow rate; and then, eq.~(\ref{rmag2}) predicts that 
\begin{equation}
r_{mag,min}\propto |\dt{\Omega}_S|^{-1/3}\, .
\label{r_omg}
\end{equation}
As was expected, a higher observed $|\dt{\Omega}_S|$ value advertises a more compressed magnetosphere, but the dependence of $r_{mag,min}$ on $|\dt{\Omega}_S|$ is very weak indeed: when the latter increases by one order of magnitude, the former is barely halved. More importantly, when units are restored in eq.~(\ref{r_omg}) we can write it in a compact form, viz.
\begin{equation}
\mu_*^2 = |\dt{\cal L}_p| (r_{mag,min})^3 \, ,
\label{mu2}
\end{equation}
where $\mu_* \equiv B_* R_*^3$ and $\dt{\cal L}_p\equiv I_* \dt{\Omega}_S$ at maximum compression ($L_X = L_{max}$). Since $\dt{\cal L}_p$ is effectively the torque applied to the pulsar by the penetrating disk, eq.~(\ref{mu2}) describes the dependence of the fully compressed magnetospheric radius on the applied disk torque at maximum luminosity. 

Eq.~(\ref{mu2}) shows that the torque $|\dt{\cal L}_p|$ on the pulsar depends on two (unobservable) parameters, $\mu_*$ and $r_{mag,min}$. Using eq.~(\ref{B2}) and eq.~(\ref{LX}), it can be recast in the form\footnote{With units restored, eq.~(\ref{torque}) reads 
$$|\dt{\cal L}_p| = \left(\eta/2\right)^{-1}\left(L_{max}/\Omega_{K*}\right) \left(r_{mag,min}/R_*\right)^{1/2}\, ,
$$
where $\eta = 1/2$, $\Omega_{K*}\equiv V_{K*}/R_*$, and $V_{K*}$ is given by eq.~(\ref{vk}).}
\begin{equation}
|\dt{\cal L}_p| \propto L_{max}\, (r_{mag,min})^{1/2} \, .
\label{torque}
\end{equation}
This proportion may explain several controversial results obtained in the late 90s for persistent X-ray sources \citep{cha97a,cha97b,nel97,bil97} that were obviously in conflict with standard accretion theory. In these investigations, the authors assumed that the torque depends solely on $L_{max}$ and they found a strong anticorrelation between $|\dt{\cal L}_p|$ and $L_{max}$ for nearly all type I and II outbursts of the persistent X-ray sources they studied. Eq.~(\ref{torque}) shows that no such conclusion can be drawn from the data before $r_{mag,min}$ is also taken into account. Nevertheless, these studies found time lags in $\dt{\cal L}_p$ relative to $L_{max}$ indicating that the magnetospheres were expanding at maximum luminosities, which implies that the tag of war between inflow and magnetic pressure may be {\it hysteretic}. Fig.~\ref{fig1} also shows this effect, although it is hard to delineate it from so few data points in the $r_{mag}/r_{co}$ plot ($r_{mag}$ expands as $L_X$ goes through its maximum). We hope that, if they are real, the hysteretic loops during the tag of war will emerge more clearly in future studies of HMXB sources for which more frequent temporal observations will become available.

\subsubsection{Pressure Equilibrium}

Eq.~(\ref{B2}) was derived under the assumption of pressure balance at the interface $r_{mag,min}=R_d$ and for $L_X=L_{max}$, when the disk has fully compressed the magnetosphere ($\dt{P_S}<0$). Pressure equilibrium at $R_d$ also implies a radial stalemate between velocities, viz.
\begin{equation}
V_A(R_d) = |V_R(R_d)|\, ,
\label{balance}
\end{equation}
where $V_A(R_d)$ is the Alfv\'en speed and $V_R(R_d)$ is the radial inflow speed at the interface. Indeed, eq.~(\ref{B2}) is equivalent to eq.~(\ref{balance}). Equivalence is obtained using eq.~(\ref{B2}) with the stellar magnetic field replaced by 
$$
B_* = B(R_d) \left(\frac{R_d}{R_*} \right)^3 \, ,
$$ 
and the mass inflow rate replaced by 
$$
\dt{\cal M}_{max} = 4\pi R_d^2\, \rho_{max}\, |V_R(R_d)| \, ,
$$ 
where $\rho_{max}$ is the mass density at the inner edge of the disk such that the Alfv\'en spped at $R_d$ is defined by
\begin{equation}
V_A(R_d) \equiv \frac{B(R_d)}{\sqrt{4\pi\rho_{max}}}\, . 
\end{equation}

\subsubsection{Mass Inflow Rate from X-ray Luminosity}

We note that eqs.~(\ref{rmag2}), (\ref{B2}), and~(\ref{balance}) do not depend on the efficiency factor $\eta/2$. This is because the equations were written in terms of $\dt{\cal M}$ which is not an observable quantity. When one proceeds to derive $\dt{\cal M}$ from the observed X-ray luminosity $L_{X}$, then
\begin{equation}
L_{X} = \frac{\eta}{2}\left(\frac{G M_* |\dt{\cal M}|}{R_*}\right) \, ,
\label{LX}
\end{equation}
and a value of $\eta$ has to be adopted. We use $\eta = 1/2$ for neutron stars and $\eta=1$ for black holes \citep[see][]{chr17b}. This equation assumes that only one half of the accretion power can be radiated away, the rest will be thermalized based on virial arguments. Furthermore, we believe that the inner disk is advection dominated and geometrically thick, not the razor-thin type described by \cite{sha73}. Additional losses (e.g., mass loss) to the emitted power are described by the efficiency parameter $\eta \leq 1$.

\subsubsection{Comparison to Standard Accretion Theory}

Finally, we address one remaining question: How do the above equations compare to the equations of standard accretion theory? The answer is that the above equations cannot be derived in the context of standard accretion because of our assumption that the disk will compress the magnetosphere and then 
$$
R_d = r_{mag} \, << \, r_{co}\, ,
$$ 
a state of vigorous accretion. On the other hand, if one were to impose hypothetically  the condition $r_{mag} = r_{co}$ to our equations, then they reduce correctly to the standard equations for the state of {\it minimum accretion} (the propeller line). Thus, standard accretion theory predicts the correct value of $B_*$ only when $L_{min}$ and $\dt{P_S}$ are measured on the propeller line and only when $L_{min}$ is certainly the propeller-line luminosity. But $\dt{P_S}$ is usually measured at maximum luminosity (ULX sources) or in the long term \citep{yang17}. Furthermore, there are always uncertainties as to whether the observed $L_{min}$ is indeed the true propeller-line luminosity. Therefore, our equations provide a firm method of determining $r_{mag}$ as a function of $L_X$ and $B_*$ at maximum luminosity, a method that does not at all suffer from such perilous uncertainties occurring in standard accretion theory.

\subsection{Retrograde Accretion Disks}

In retrograde accretion disks, in which the flow counterrotates relative to the pulsar, there is no corotation radius, thus conventional accretion cannot take place. Yet, pulsar spindown is as common as spinup and of the same $|\dt{P_S}|$ magnitude \citep{chr17a}. Many researchers have objected to the process of retrograde accretion because it does not fit into standard theory. But our observational results of pulsars in the Small Magellanic Cloud are conclusive: there is absolutely no statistical difference between populations of spinning up and spinning down pulsars. In fact, except for the sign of $\dt{P_S}$, Be/X-ray binary pulsars all appear to be very similar, with the Be stars supplying matter to form accretion disks at least once in every orbit and the neutron stars exhibiting regular type I outbursts (and occasionally more powerful type II outbursts) \citep{coe10,reig11}.

The statistical results can be understood in the context of the new mechanism as follows. A retrograde disk counterrotates at all radii, therefore it always applies a retarding torque to the magnetosphere. In fact, this torque becomes stronger interior to the counterrotation radius $R_{crr}$ defined as the radius where the disk's rotation period equals the pulsar's spin period, viz.
$$
P_d(R_{crr}) = P_S\, .
$$
But loading of the field lines proves to be problematic at radii $R\lsim R_{crr}$ because of the oppositely directed velocity and angular momentum vectors (it is like trying to jump on a merry-go-round while running around it at the opposite speed). 

In such a case, the retrograde accretion disk could only continue to slowly spin down the pulsar forever, but this scenario is contradicted by observations of the slowest spinning pulsars such as SXP1062 and SXP1323 which reversed their $\dt{P_S}$ signs in a matter of days. We must then conclude that the observed state transitions from spindown to spinup at slow spins occur exclusively in prograde disks when the magnetospheres are pushed back inside corotation by enhanced matter inflow. 

Our conclusion justifies and supports all those who doubted retrograde accretion in the past on the basis of standard accretion theory. Prograde disks with their magnetospheres expanding or contracting in response to changes in $\dt{\cal M}$ also explain why spinning down pulsars are as common as spinning up pulsars \citep{chr17a} and the many reversals occurring between the two types in the short term \citep{yang17}. 

\section{Conclusions}\label{conc}

The observed enormous spin period rates $\dt{P_S}$ observed in the X-ray sources listed in Tables~\ref{table1} and~\ref{table2} cannot be explained by standard spin-equilibrium theory. We are forced to consider new accretion mechanisms and evolutionary paths in which enormous $\dt{P_S}$ values are produced, not near the surfaces of the pulsars, but in their extended magnetospheres. The magnetic fields can then apply torques to the compact objects from a distance; there is no requirement that matter along with its angular momentum should descend all the way down to the polar caps before a large spinup can be observed. 

Indeed, the magnetic field lines are capable of spinning up or down the pulsars depending on whether their magnetospheres are crushed by the weight of increased $\dt{\cal M}$ or they push outward beyond corotation when inflow begins to taper off, respectively. A mechanism such as this can also explain intermittent variability, enormous abrupt reversals in the sign of $\dt{P_S}$ (Table~\ref{table4}), and argues against retrograde accretion that would have to take place in the absence of a corotation radius, an impossible feat. The Magellanic pulsar SXP1062 and the ULX pulsar in NGC300 emerge as prototypical examples of X-ray sources that exhibit new phenomena, such as enormous $\dt{P_S}$ rates and, for the former, reversals in the sign of $\dt{P_S}$ within just $\sim$10 days as the magnetosphere competes against the weight of inflowing matter. 

We are at the cusp of great discoveries in binary pulsar evolution, this is why the X-ray outbursts of sources such as those listed in Tables~\ref{table2} and~\ref{table3} must be observed on a daily basis for their extreme state transitions to be recorded in detail \citep[as was done by][]{cha97a,bil97}. Differences in their evolutionary paths should reflect the different magnitudes of their magnetic fields: SXP1062 appears to harbor a strong magnetic field, about 100 times stronger than that of NGC300 ULX1, and this is clearly due to its $r_{mag,min}$ being larger by a factor of 118 (Table~\ref{table3}). The properties of the ULX1 source are however very uncertain because they depend strongly on the amount of beaming of the X-ray radiation. Meanwhile, the third pulsar, SXP1323, should continue to spin up in the near future as its weak magnetic field (see Table~\ref{table3}) was crushed by enhanced inflow in 2005, making it very hard for it to bounce back and expand back out toward corotation until $\dt{\cal M}$ tapers off considerably.

\section*{Acknowledgments}
We are obliged to an anonymous referee for an extremely thorough review of the paper and for the many suggestions that improved considerably the presentation of our ideas. DMC and SGTL were supported by NASA grant NNX14-AF77G. DK was supported by a NASA ADAP grant.

\label{lastpage}

\end{document}